# Identification of false positive double-lined spectroscopic binaries in LAMOST-MRS data due to moonlight contamination

Zhoulin Wang,[1,2,3] Mikhail Kovalev ,[1,3,4,5] Jianping Xiong,[1,3,4] Yanjun Guo[1,3,4] and Xuefei Chen[1,3,4]★

[1] *Yunnan Observatories, Chinese Academy of Sciences, 396 YangFangWang, Guandu District, Kunming 650216, People's Republic of China*
[2] *International Centre of Supernovae, Yunnan Key Laboratory, Kunming 650216, People's Republic of China*
[3] *School of Astronomy and Space Science, University of the Chinese Academy of Sciences, Beijing 100049, People's Republic of China*
[4] *Key Laboratory for Structure and Evolution of Celestial Objects, Chinese Academy of Sciences, P.O. Box 110, Kunming 650216, People's Republic of China*
[5] *Center for Astronomical Mega-Science, Chinese Academy of Sciences, 20A Datun Road, Chaoyang District, Beijing 100012, People's Republic of China*

## ABSTRACT

We present a method for identifying false positive double-lined spectroscopic binary (SB2) candidates by analysing medium-resolution survey spectra from the Large Sky Area Multi-Object Fiber Spectroscopic Telescope (LAMOST) DR10. Specifically, we focus on spectra contaminated by moonlight, which exhibit near-zero radial velocity (RV) and solar-like spectral lines from the secondary component. By applying strict constraints on the contamination parameters and fitting the contaminated spectra, we ultimately confirmed that 126 false positive binaries are single stars contaminated by moonlight. Additionally, we identify several key factors contributing to moonlight contamination: the lunar phase during observation, the *G*-band magnitude of the star, and the angular distance between the star and the moon. Notably, artificial satellites in low-orbital can also introduce contamination from solar-like spectral components, but they typically display significantly higher transverse velocities. In a follow-up study, we will expand our analysis to identify additional false positive SB2 systems and systematically classify them according to their contamination sources.

**Key words:** light pollution – techniques : spectroscopic – binaries : spectroscopic.

## 1 INTRODUCTION

Binary star systems are prevalent throughout the universe. Over 60 per cent of Milky Way stars are part of binary or multiple systems (Abt 1983). In modern astrophysics, binary systems are considered fundamental due to their extensive connections with various research areas, including stellar formation, stellar physics, galaxies, and cosmology [e.g. the formation of gravitational waves (Schneider et al. 2001); the black hole binary research (Liu et al. 2019); binary formation, evolution and interactions (Hilditch 2001; Han et al. 2020)].

In recent years, researchers have extensively conducted statistical investigations of binary systems. Due to the lack of a sufficient number of binary star samples, the statistical distributions of masses, orbital periods, mass ratios, and eccentricities remain poorly understood (Duquennoy & Mayor 1991; Halbwachs et al. 2003; Raghavan et al. 2010; Moe & Di Stefano 2017; Guo et al. 2022). Fortunately, large-scale, multi-epoch photometric, spectroscopic, and astrometric surveys such as the All Sky Automatic Survey for Supernovae (ASAS-SN), the Zwicky Transient Facility (ZTF), KEPLER, *TESS*, the Sloan Digital Sky Survey (SDSS), LAMOST, *Gaia*-ESO Survey, the GALactic Archaeology with HERMES (GALAH) and the Apache Point Observatory Galactic Evolution Experiment (APOGEE) have been conducted. These surveys have allowed researchers to identify a substantial number of binaries and investigate their properties.

Double-lined spectroscopic binaries (SB2s) are intriguing objects, as they allow us to provide a measurement for the absolute individual mass and radius of each star with accuracies better than <1 per cent (Ge et al. 2010). They are commonly detected through the analysis of double peaks in the cross-correlation function (Matijevič et al. 2011; Merle et al. 2017; Li et al. 2021; Kounkel et al. 2021). Moreover, recent advancements in machine learning techniques have facilitated the identification of SB2s (Traven et al. 2017, 2020; Zhang et al. 2022). Additionally, spectral fitting or disentangling methods, which employ data-driven spectral models to separate individual components of the binary system, are effective for discovering SB2s (El-Badry et al. 2018a, b; Kovalev, Chen & Han 2022; Kovalev et al. 2024).

Numerous catalogues of spectroscopic multiple-star systems have been published: Pourbaix et al. (2004) compiled the ninth catalogue of SB orbits, which includes over 4004 SBs, with approximately one-third classified as SB2. The Geneva–Copenhagen Survey presents 3223 SBs from 16 682 nearby F and G dwarf stars (Nordström et al. 2004; Holmberg, Nordström & Andersen 2009). 123 SBs have been found in the Radial Velocity Experiment survey (Matijevič et al. 2010). Over the past decade, spectroscopic surveys have significantly expanded, facilitating the detection of binaries. Merle et al. (2017) recognized 342 SBs, 11 STs, and 1 quadruple-line candidate from the *Gaia*-ESO survey. The APOGEE survey discovered over 3000 binaries (Fernandez et al. 2017; El-Badry et al. 2018b; Kounkel et al. 2021). In the GALAH survey, 12 760 F, G, and K SBs were identified, as reported by Traven et al. (2020). Additionally, 3133, 2198, and 12 426 SB2 candidates were detected in the LAMOST survey by Li et al. (2021), Zhang et al. (2022), and Kovalev et al. (2024), respectively.

★ E-mail: cxf@ynao.ac.cn

During spectroscopic observations, contamination can cause a single star to be misidentified as a double-lined spectroscopic binary, with the contaminant typically manifesting as a secondary component. For instance, the flyby of a low-orbit satellite can result in a fake spectroscopic binary during observations (Bassa, Hainaut & Galadí-Enríquez 2022). Kovalev et al. (2022, 2024) identified numerous double-lined spectroscopic binary candidates in LAMOST-MRS, some of which may be false positives. We present results aimed at detecting false positive SB2s caused by moonlight, using time-domain data from LAMOST DR10.

The paper is organized as follows: In Sections 2 and 3, we describe the data and methods. Section 4 presents our results. In Section 5, we summarize the paper.

## 2 DATA

LAMOST is a 4-m reflective Schmidt telescope equipped with 4000 fibers installed on its 5° field of view focal plane (Cui et al. 2012; Zhao et al. 2012). This survey commenced in 2011 and has released spectra for over 10 million stars (Liu et al. 2020). Since 2018, LAMOST has conducted a medium-resolution spectroscopic survey (MRS) during bright/grey nights, achieving a resolution of $R \sim 7500$. The wavelength range spans $495 \sim 535$nm for the blue band and $630 \sim 680$nm for the red band, targeting stars with magnitudes $9.0 \leq G_{\rm mag} \leq 15.0$ mag.

Some spectra may be affected by moonlight, resulting in residual background glow (Liu et al. 2020). The LAMOST fiber-optic system is divided into 16 regions, each containing 250 fibers connected to a spectrograph. In each region, approximately 20 fibers are dedicated to sampling the sky background. The sky subtraction process involves averaging the spectra from these sky fibers and uniformly subtracting that average from all regional fibers. Therefore, when the moon is bright, there is a significant brightness gradient across the sky. The efficiency of each fiber is different, for fibers that are more severely affected, even after sky subtraction, the moonlight component may still remain. Therefore, moonlight contamination is unavoidable to some extent. When the MRS mode operates during bright nights, and the moon phase is nearly full Moon, stars polluted by moonlight may be misidentified as binaries. Therefore, the SB2 candidates identified by Kovalev et al. (2024) based on LAMOST-MRS data are highly likely to be affected by moonlight contamination. For these binaries, we download all available time-domain data in data release 10 from www.lamost.org/dr10/v0/, resulting in 42 599 spectra of 12 353 stars, converting the heliocentric wavelength scale from vacuum to air using PYASTRONOMY (Czesla et al. 2019). All spectra are stacked within a single night, and a signal-to-noise threshold of $S/N \geq 20$ pix$^{-1}$ is applied to select data for analysis.

## 3 METHODS

### 3.1 Spectroscopic models

Before discussing the search for potential false positive binaries, we first summarize the methodologies introduced by Kovalev (Kovalev et al. 2022, 2024). They proposed new spectroscopic models for detecting double-lined spectroscopic binaries:

The normalized binary spectral model ($f_{\lambda,\rm binary}$) is generated as a sum of the two Doppler-shifted normalized single-star spectral models, $f_{\lambda,1}$ represents the primary star, $f_{\lambda,2}$ represents the secondary star. Each $f_{\lambda,\rm i}$ is scaled according to the difference in luminosity, which is a function of the effective temperature ($T_{\rm eff}$) and stellar size. Assuming both components to be spherical, represented by the following equation:

$$f_{\lambda,\,\rm binary} = \frac{f_{\lambda,2} + k_\lambda f_{\lambda,1}}{1 + k_\lambda},\, k_\lambda = \frac{B_\lambda\left(T_{\rm eff\,,1}\right) R_1^2}{B_\lambda\left(T_{\rm eff\,,2}\right) R_2^2}, \quad (1)$$

where $k_\lambda$ is the luminosity ratio per wavelength unit, $B_\lambda$ is the blackbody radiation. The binary model spectrum is later multiplied by the normalization function, a linear combination of the first four Chebyshev polynomials (Kovalev 2019), and is considered for the blue and red arms separately. The result is compared with the observed spectra by `scipy.optimize.curve_fit` function, which provides the optimal spectral parameters and RVs for each component; the stellar radii ratio $k_{\rm R} = R_1/R_2$ and two sets of four coefficients of the Chebyshev polynomials.

Each spectrum is fitted with the single-star spectral model, and the solution is marked as $f_{\lambda,0}$. They compute the discrepancy in the reduced $\chi^2$ between the two models, as well as the improvement factor ($f_{\rm imp}$), which is calculated using equation (2), similar to the method described in (El-Badry et al. 2018b). This improvement factor estimates the absolute value difference between two fits and, weights it by the difference between the two solutions:

$$f_{\rm imp} = \frac{\sum_{\lambda=\lambda_{\rm min}}^{\lambda=\lambda_{\rm max}}\left[\left(\left|f_{\lambda,0} - f_\lambda\right| - \left|f_{\lambda,\rm binary} - f_\lambda\right|\right)/\sigma_\lambda\right]}{\sum_{\lambda=\lambda_{\rm min}}^{\lambda=\lambda_{\rm max}}\left[\left|f_{\lambda,0} - f_{\lambda,\rm binary}\right|/\sigma_\lambda\right]}, \quad (2)$$

where $f_\lambda$ and $\sigma_\lambda$ represent the observed flux and corresponding uncertainty per wavelength respectively. $f_{\lambda,0}$ and $f_{\lambda,\rm binary}$ are fitted from the single-star spectral model and the binary spectral model and summed over all wavelength pixels. As for $\lambda_{\rm min}$ and $\lambda_{\rm max}$, they denote the range of observational wavelengths. In general, the larger value of $f_{\rm imp}$ indicates a higher likelihood that the star is a binary system.

If the secondary component arises from the presence of a companion star, it suggests a genuine binary system. Conversely, if the secondary component results from lunar illumination, it indicates a false positive SB2, which is the subject of our investigation.

### 3.2 Remove eclipsing binaries

We need to identify false positive SB2 due to moonlight contamination. Therefore, it is necessary to exclude EBs before the analysis. We identified 24 matches with the KEPLER eclipsing binary catalogue (Slawson et al. 2011), and 43 matches through cross-matching with the *TESS* eclipsing binary catalogues (IJspeert et al. 2021; Prša et al. 2022). Additionally, 272 objects were found to correspond with the ZTF eclipsing binary catalogue (Chen et al. 2020) and 370 matches were identified in the ASAS-SN eclipsing binary catalogues (Rowan et al. 2022, 2023a, b). After removing duplicates, we matched a total of 624 eclipsing binaries. Excluding these eclipsing binaries, we retained 11 729 samples for further analysis.

### 3.3 Selection of contaminated stars

Contamination events in spectra can be classified into two types: those where contamination is detected only once across multiple time-domain observations, and those where contamination appears repeatedly. Analysing stars with multiple contaminated spectra is more complex; therefore, we prioritized stars that exhibit a single composite spectrum and excluded cases with repeated contamination. Moreover, in stellar systems with multiple observations, if only one spectrum displays binary characteristics, it is more likely attributable to contamination. Contamination events in spectra can be categorized into two types: those where contamination occurs only once during multiple time-domain observations, and those where contamination occurs multiple times across the observations. However, analysing stars with multiple contaminated spectra is more complex. Therefore, we prioritized studying stars with only one composite spectrum and excluded cases with multiple contaminated spectra. In stellar systems with multiple observations, where only one spectrum displays binary characteristics, that spectrum is more likely to be contaminated. When a single star is contaminated by moonlight, it may display two distinct components with differing RV. Moonlight contamination typically presents a solar-like spectrum. Additionally, the Moon, as a nearby celestial body, has a very slow rotational velocity, meaning that the $V\sin i$ value of the contamination is expected to be low.

**Table 1.** Initial selection criteria.

| To identify potentially contaminated stars in SB2 candidates. |
|---|
| $N_{obs} > 4$ |
| Only one spectrum is composite |
| $\min(V\sin i_1, V\sin i_2) < 30\,\mathrm{km\,s^{-1}}$ |
| $\min(\lvert T_{\mathrm{eff1}} - 5777\rvert, \lvert T_{\mathrm{eff2}} - 5777\rvert) < 700$ K |

The estimation errors in $V\sin i$ and $T_{\mathrm{eff}}$ reported by Kovalev et al. (2022) are approximately $\sim 13\,\mathrm{km\,s^{-1}}$ and $\sim 192$ K for the primary star, and up to $\sim 60\,\mathrm{km\,s^{-1}}$ and $\sim 891$ K for the secondary star. Considering the precision of the estimated parameters, we have established a set of initial selection criteria to identify potentially contaminated stars among the SB2 candidates, as detailed in Table 1. These criteria consist of four conditions: the first two constraints are designed to screen out false positive SB2 candidates. The parameter $N_{obs}$ represents the number of observations. The remaining two are aimed at identifying objects affected by moonlight.

Based on these criteria, we identified 374 objects, some of which are true binaries, while others are single stars contaminated by scattered moonlight. To ensure that no genuine binary systems were misclassified as contaminated single stars, we conducted a meticulous visual inspection of all 374 stars. For each star, we examined all time-domain spectra, as well as the fitting results from both single and binary spectral models (Kovalev et al. 2024), along with key parameters such as radial velocity, atmospheric parameters, rotational velocity, improvement factor, and moon phase.

For a star with '$n$' times observations, correspond to '$n$' spectra. Our criteria dictate that $n-1$ spectra must exhibit single-star characteristics. We categorize these $n-1$ spectra as Type-1, while the spectrum showing two components is classified as Type-2. The Type-1 spectra should exhibit similar $RV_0$ values. For the Type-2 spectrum, the primary component should resemble the Type-1 spectra, and $RV_1$ should be similar to $RV_0$ of the Type-1 spectra. The secondary component is expected to exhibit solar-like spectral lines, with $RV_2 + RV_h \sim 0$, where $RV_h$ represents the heliocentric correction to the radial velocity. The LAMOST spectral data is employed in the heliocentric reference system, it cannot avoid the effects of the Moon's rotation and orbit around the Earth. To avoid this issue, we transform the data to a geocentric coordinate system by taking $RV_h$ into account. We will elaborate on the details and provide an example in the subsequent section.

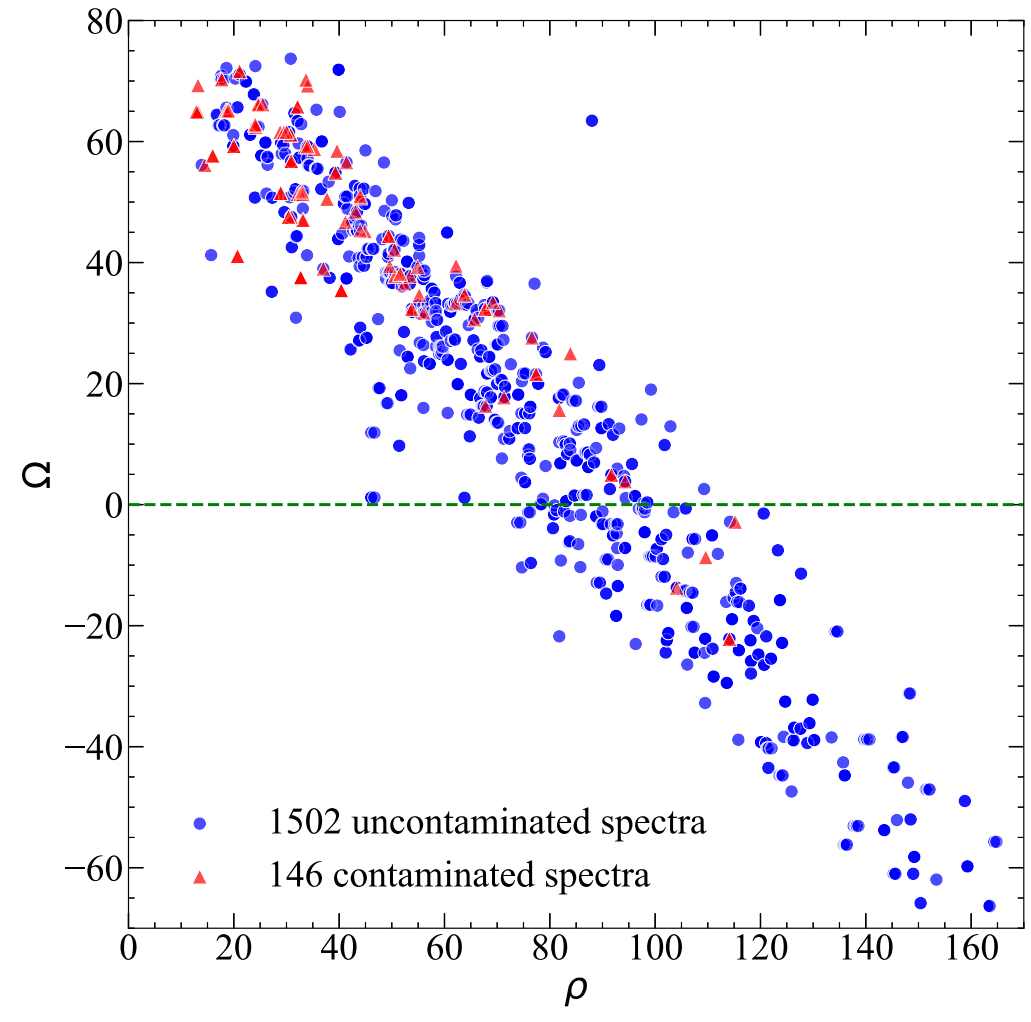

**Figure 1.** The Moon's altitude and angular distance of Type-1 and Type-2 spectra are shown. A green dashed line divides the data into two sections, with the Moon altitude being negative below this line.

**Table 2.** Five spectra were observed at a negative moon altitude.

| Designation | RA, ° | Dec, ° | MJD, d | $\rho$, ° | $\Omega$, ° |
|---|---|---|---|---|---|
| J102842.82+391742.0 | 157.178 | 39.295 | 58951 | 114.1 | −22.18 |
| J103011.47+392846.0 | 157.548 | 39.479 | 58951 | 114.1 | −22.18 |
| J085631.66+241914.3 | 134.132 | 24.321 | 59179 | 104.1 | −13.74 |
| J125506.74+530011.2 | 193.778 | 53.003 | 59663 | 109.6 | −8.69 |
| J005931.70+050839.4 | 14.882 | 5.144 | 58477 | 115.2 | −2.83 |

## 4 RESULTS

After a meticulous visual inspection, we identified 228 stars exhibiting strong evidence of binarity and 146 stars as potentially contaminated, with the latter corresponding to a total of 1650 spectra. 1504 uncontaminated spectra were classified as Type-1, and 146 contaminated spectra were classified as Type-2. In general, the likelihood of contamination increases as the Moon approaches the star. To investigate this, we cross-match with the LAMOST-MRS Parameter Catalogue and obtained $\rho$ (angular distance between object and moon). It is worth noting that two Type-1 spectra lack $\rho$ measurements. Additionally, we utilize `astropy.coordinates` and `astropy.time` to calculate the Moon's altitude, denoted as $\Omega$. Fig. 1 presents the distribution of $\rho$ and $\Omega$ for both Type-1 and Type-2 spectra, revealing a negative correlation between these two parameters.

For Type-2 spectra, the $\Omega$ values are generally greater than 0. Since moonlight contamination is not expected when the Moon's altitude is below the horizon, we excluded five potentially contaminated spectra with negative moon altitudes in further analysis. However, Fig. 1 only displays four targets. Therefore, we present detailed information for all five observations in Table 2. We found that J102842.82+391742.0 and J103011.47+392846.0 have the same Modified Julian Date (MJD), resulting in identical $\Omega$ values. Furthermore, their Right Ascension (RA) and Declination (Dec) are similar, indicating nearly identical positions in the sky. Consequently, their $\rho$ values are also identical, causing them to overlap in Fig. 1.

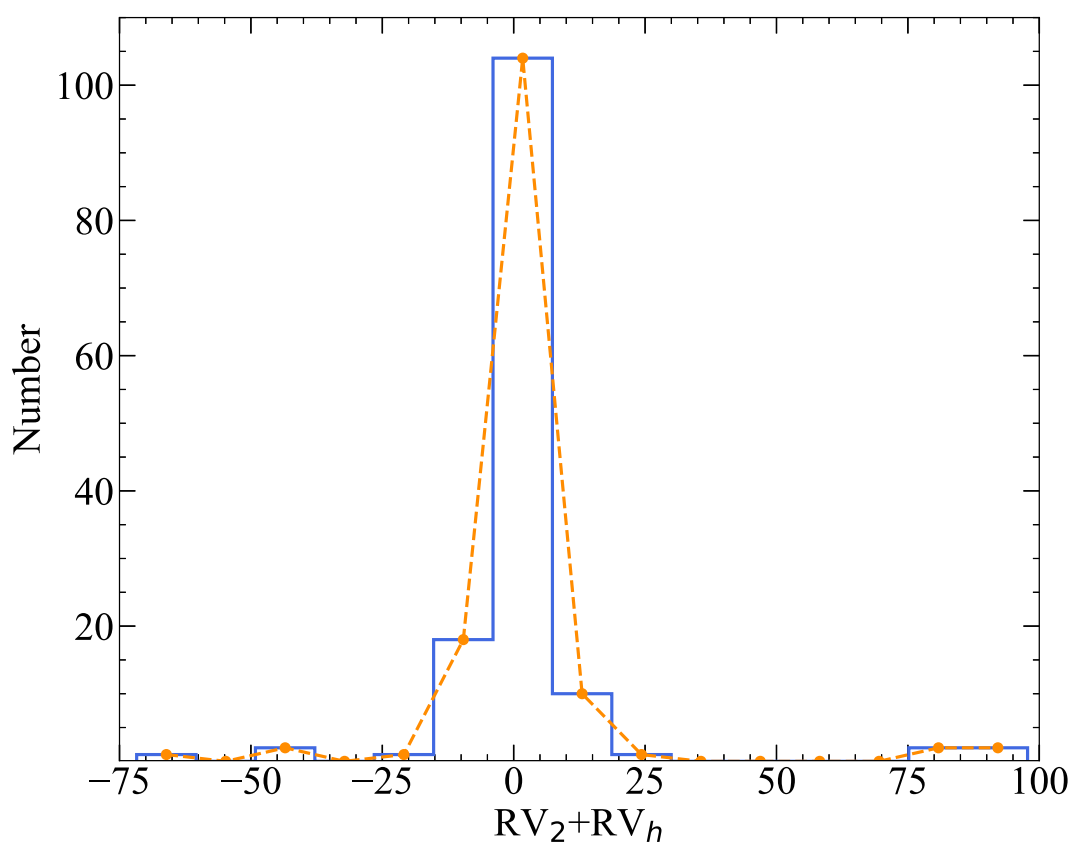

**Figure 2.** The distribution of $RV_2 + RV_{\rm h}$ for the potential contaminated component.

For the remaining 141 contaminated spectra, we utilized the binary spectra model to fit and determine the radial velocities of both components, with $RV_2$ corresponding to the contaminated components. Additionally, $RV_2 + RV_{\rm h}$ represents the radial velocity of the secondary component relative to Earth. The distribution of $RV_2 + RV_{\rm h}$ is shown in Fig. 2, most spectra exhibit small values of $RV_2 + RV_{\rm h}$. In general, $RV_2$ is approximately equal to $-RV_{\rm h}$ if the secondary component is moonlight pollution. Taking into account the precision of both $RV_2$ and $RV_{\rm h}$, we established a selection criterion of $|RV_2 + RV_{\rm h}| < 10$ km $\cdot$ s$^{-1}$ and identified 126 contaminated candidates for further analysis. In the next two subsections, we detailed the process of identifying contaminated spectra and displayed the statistical characteristics of these 126 stars. We found that 15 spectra did not satisfy this criterion. As a result, these spectra were excluded from subsequent analysis, with a detailed discussion provided in Appendix A.

### 4.1 An example of detecting false positive SB2

We use J114138.14+364153.6 as an example to illustrate the process of identifying a contaminated spectrum. As reported by Kovalev et al. (2024), J114138.14+364153.6 has been classified as a double-lined spectroscopic binary candidate. We derived parameters for this star, using both binary and single-star spectral models. This object was observed nine times in the LAMOST-MRS survey. Only one contaminated spectrum (blue line) shows two distinct components (see Fig. 3), particularly noticeable in the H $\alpha$ and Mgb lines. For this spectrum, three times 20-min exposure spectra are presented in Fig. 4. The exposure began at 00:43 and ended at 01:34, corresponding to the local time (UTC + 8). Given the low-orbital artificial satellite's high transverse velocity, any contamination from the satellite would be short, affecting only one short 20-min exposure. In Fig. 4, each spectrum shows two components, strongly indicating that the contamination is due to the Moon rather than a satellite.

We conducted a detailed inspection of the other eight spectra for J114138.14+364153.6 and found that they are nearly identical, exhibiting typical single-star features. For a more in-depth analysis, we provide the values of $RV_0$, $RV_1$, and $RV_2$ in Table 3. Except for MJD = 58918.721, the other spectra display nearly identical $RV_0$ values. Moreover, in these eight spectra, both $RV_1$ and $RV_2$ closely resemble $RV_0$.

Furthermore, the contaminated spectrum shows two distinct components: $RV_1$ value closely matches the $RV_0$ values of the eight uncontaminated spectra. While the secondary component exhibits a significantly smaller $RV_2$ (see Table 3). After applying geocentric correction, we found $RV_2 + RV_{\rm h} \sim 2.21$ km $\cdot$ s$^{-1}$, suggesting that the secondary component is most likely due to moonlight contamination.

To further verify the moonlight contamination, we attempted to fit the polluted spectrum, using a solar-like spectrum and an uncontaminated spectrum. We selected a spectrum with the minimum $f_{\rm imp}$ value as the uncontaminated spectrum (MJD: 58851.903). The solar spectrum was obtained from https://bass2000.obspm.fr/solar_spect.php. We normalized it, and then applied Gaussian instrumental broadening using `PyAstronomy` to match the LAMOST resolution of $R = 7500$.

The spectra of J114138.14+364153.6 from LAMOST were recorded in a consistent observational coordinate system, while the solar spectrum was treated as a non-shifted spectrum. Consequently, we employ the value of $RV_2 = -1.94$km $\cdot$ s$^{-1}$ derived from the contaminated spectrum to correct the solar spectrum for the Doppler shift using the following equation:

$$\lambda = \lambda_0 \left(1 + \frac{RV_2}{c}\right). \tag{3}$$

In equation (3), $\lambda$ represents the Doppler-shifted wavelength, while $\lambda_0$ denotes the unshifted wavelength of the solar spectrum. $RV_2$ is the radial velocity derived from the binary spectra model for the contaminated spectrum. Similarly, we adjust the radial velocity of the uncontaminated spectrum to match $RV_2$ of the contaminated spectrum, thereby minimizing systematic errors arising from different observations.

Next, we resample the wavelengths of these spectra and utilize `scipy. optimize.curve_fit` for non-linear least squares fitting. We define a fitting parameter and fit the contaminated spectrum by combining the uncontaminated spectrum and the shifted solar spectrum, as described in equation (4).

$$f_{\lambda,{\rm c}} \propto \gamma \cdot f_{\lambda,{\rm n}} + (1-\gamma) \cdot f_{\lambda,{\rm solar}}. \tag{4}$$

In equation (4), $f_{\lambda,{\rm c}}$, $f_{\lambda,{\rm n}}$, $f_{\lambda,{\rm solar}}$ represent the normalized flux per wavelength for the contaminated spectrum, the uncontaminated spectrum, and the solar spectrum, respectively. The parameter $\gamma$ represents the contribution rate of the two components in the fitted spectrum. Since LAMOST MRS spectral data are split into blue and red arms – we introduce separate $\gamma$ parameters for the blue and red arms.

The fitting result is displayed in Fig. 5, with a contribution rate of 0.634 for the blue and 0.557 for the red arms. The slightly high value in the blue arms may indicate that this star is hotter than the Sun. The green line represents the fitted spectrum, composed of the uncontaminated and solar spectra. The uncontaminated and solar spectra, scaled by $\gamma$, are shown in red and blue, respectively. The grey line corresponds to the observed spectrum, while the magenta line indicates observational errors. Additionally, the bottom panel of Fig. 5 shows the residuals between the contaminated and fitted spectra.

One distinguishing feature of our approach is that we used observed spectra, rather than template spectra, which enhances the reliability of our fitting results. For the quantitative assessment of our fitting results, we also provided a reduced chi-squared value ($\chi^2_\nu$, see Fig. 5).

$$\chi^2_\nu = \frac{1}{n-m} \sum_{i=1}^{n} \left( \frac{F^i_{\rm obs} - F^i_{\rm fit}}{\sigma^{i\,2}_{\rm e}} \right)^2. \tag{5}$$

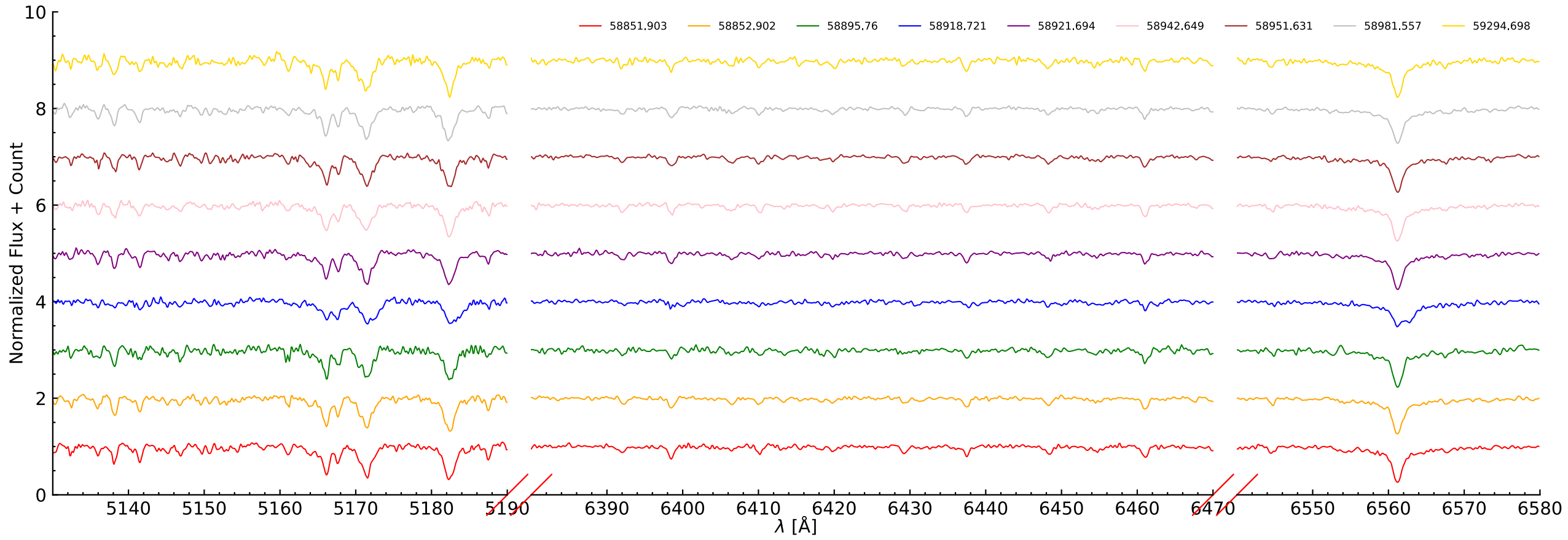

**Figure 3.** The time-domain co-added spectra of J114138.14+364153.6 are sorted by MJD and labelled at the top of the image. The spectrum of MJD:58918.721 is contaminated, while the remaining spectra display the same characteristics as a typical single star.

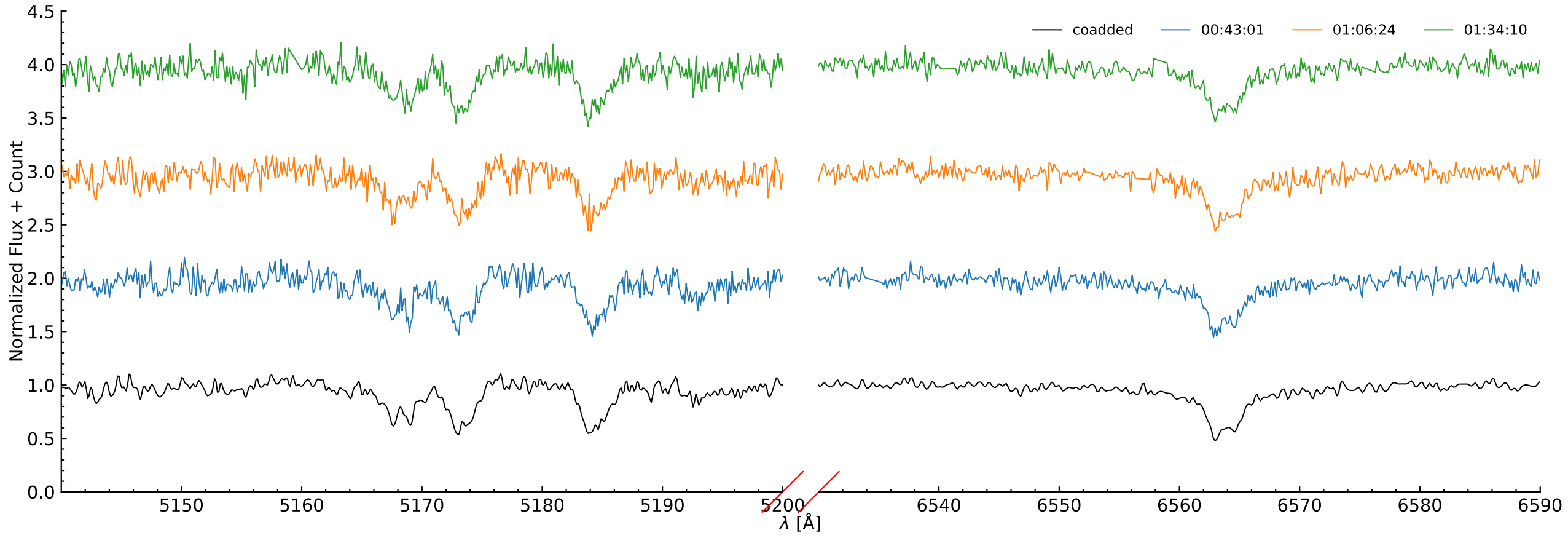

**Figure 4.** The contaminated spectrum of J114138.14+364153.6 at MJD: 58918.721 d, displays individual 20 min exposures, with the times indicated in the legend representing the local time at which each observation began during the exposure.

**Table 3.** Fitting parameters of nine spectra of J114138.14+364153.6. We highlight the contaminated spectrum in boldface.

| MJD, d | $RV_0$, $\mathrm{km\cdot s^{-1}}$ | $RV_1$ | $RV_2$ | $RV_h$ | $f_{imp}$ |
|---|---|---|---|---|---|
| 58851.903 | −73.40 | −77.11 | −72.74 | −21.74 | −0.014 |
| 58852.902 | −72.67 | −77.58 | −72.13 | −21.51 | −0.001 |
| 58895.760 | −73.31 | −77.81 | −72.46 | −6.05 | 0.004 |
| **58918.721** | **−42.18** | **−73.29** | **−1.94** | **4.15** | **0.220** |
| 58921.694 | −72.79 | −104.18 | −72.0 | 5.40 | 0.083 |
| 58942.649 | −73.16 | −77.60 | −73.03 | 13.83 | 0.106 |
| 58951.631 | −72.54 | −109.8 | −71.83 | 16.89 | 0.033 |
| 58981.557 | −74.89 | −71.19 | −76.96 | 23.79 | 0.002 |
| 59294.698 | −74.64 | −71.48 | −76.73 | 8.73 | −0.008 |

The calculation of reduced chi-square is given in equation (5). In this equation, $n-m$ represents the degrees of freedom, where $F_{\rm obs}$, $F_{\rm fit}$, and $\sigma_{\rm e}$ denotes the flux of contaminated spectrum, the flux of fitted spectrum, and the errors, respectively. Here, $n$ is the sample size. $m$ is the number of variables, set to 2 due to the two $\gamma$ parameters. The calculated value of $\chi^2_\nu$ value is 0.593, indicating that the fitting results are highly satisfactory.

All the analyses above suggest that J114138.14+364153.6 is likely a single star affected by moonlight, despite its previous classification as a potential double-lined spectroscopic binary. The process is applied to each contaminated spectrum.

## 4.2 Statistics and analysis

We presented the statistical properties of the potentially contaminated stars and examined the correlation between moonlight contamination and lunar activity, as well as other factors.

The spectrum is also more likely to be contaminated during the full moon period. To investigate this, we compiled a distribution of the lunar phase of these 126 contaminated spectra, as shown in Fig. 6 (a). The values for all the objects fall within the range of the 10th to the 20th day. The distribution shows a prominent peak around the 15th day, corresponding to the full moon phase, where contamination frequency is highest. The numbers rapidly decrease on either side of the peak, consistent with the expected moon-related behaviour.

Distance is a key factor in assessing the likelihood of contamination: the closer the Moon is to the observed source, the greater the possibility of contamination. We present the distribution of $\rho$ values for 126 spectra in Fig. 6 (b), which shows a stepwise decline in the number of occurrences as $\rho$ increases. Even a star located at a considerable angular distance from the Moon may still be contaminated. We suggest that this contamination could result from moonlight reflected off clouds, the ground, or other environmental factors.

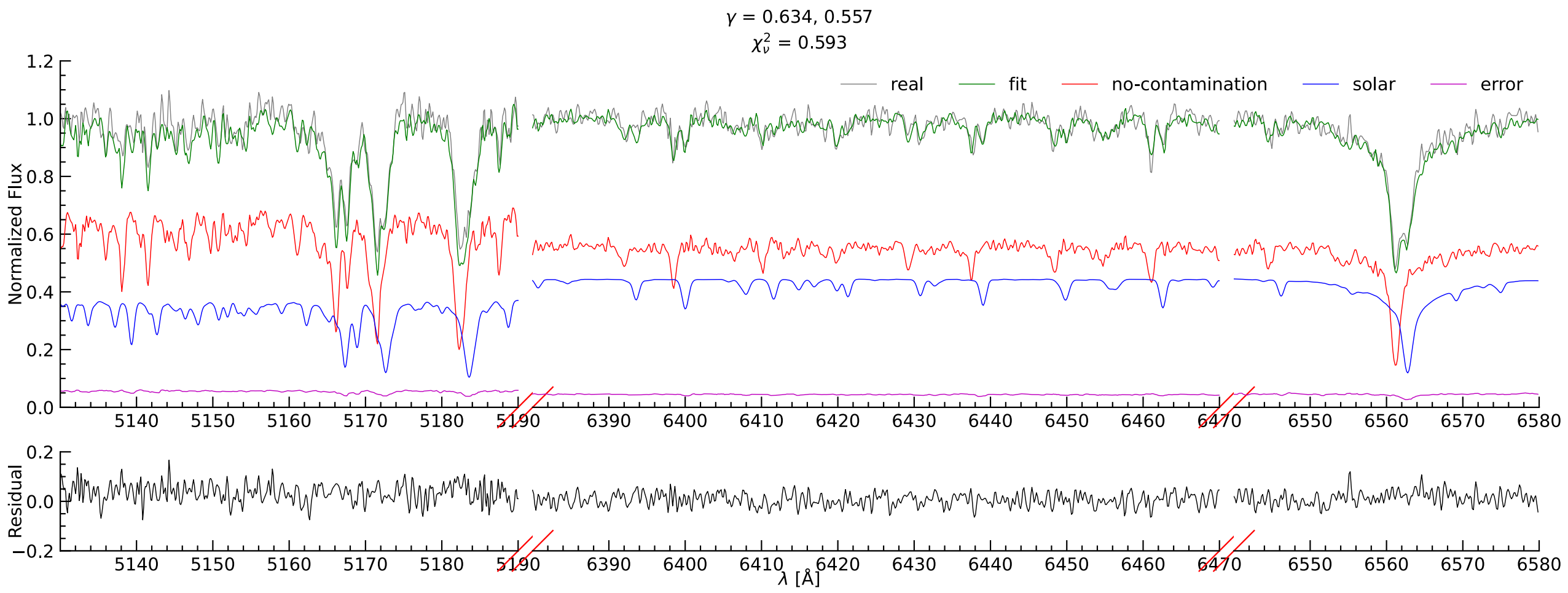

**Figure 5.** The top panel displays the fitting results, where the grey lines represent the contaminated spectrum. The green line shows the combined fit of the uncontaminated (MJD:58851.903) and solar spectra, marked red and blue, respectively (normalized flux scaled by the parameter $\gamma$). The magenta line shows the error. The bottom panel illustrates the residuals between the observed spectrum and the fitted one, calculated as $F_{\lambda,\mathrm{real}} - F_{\lambda,\mathrm{fit}}$.

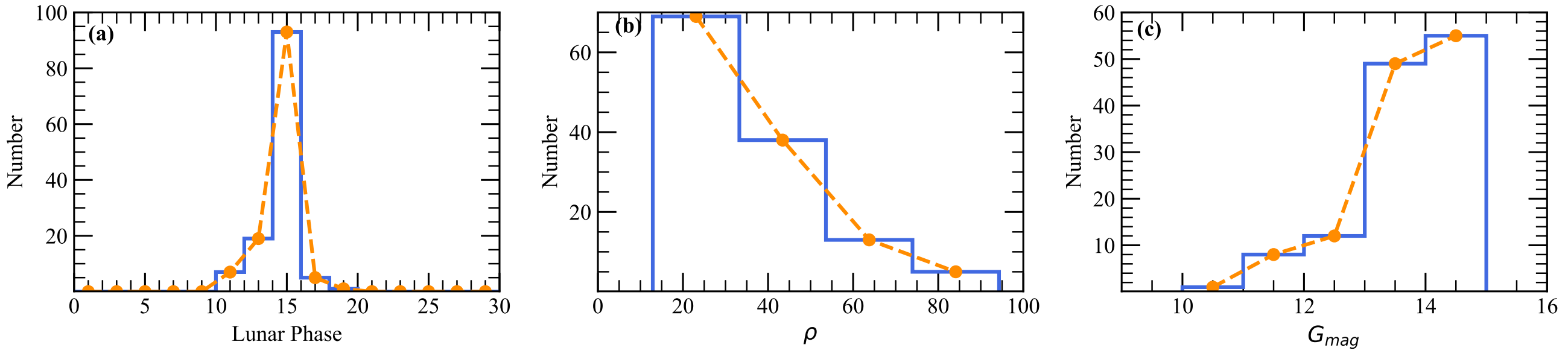

**Figure 6.** In figure (a), the lunar phases recorded in the FITS files derived from the LAMOST-MRS data are analysed. Figure (b) illustrates the distribution of $\rho$ for detected objects, showing a decrease in frequency as $\rho$ increases. Figure (c) depicts the $G_{\mathrm{mag}}$ distribution of false-positive SB2 objects, revealing that most of these objects are faint.

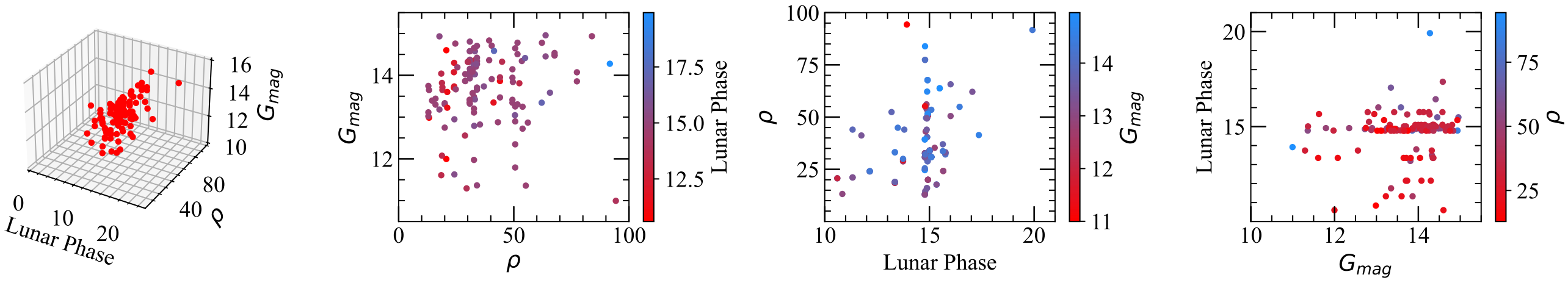

**Figure 7.** A 3D scatter plot and three 2D scatter plots with colour bars in different directional slices are created using the lunar phase, $\rho$, and $G_{\mathrm{mag}}$. Each point represents a spectrum affected by moonlight contamination.

Additionally, we contemplate that fainter stars are more susceptible to contamination. To investigate this, we derive the $G_{\mathrm{mag}}$ ($G$-band magnitude) by cross-match with *Gaia* DR3 data (Prusti et al. 2016; Vallenari et al. 2023). The $G_{\mathrm{mag}}$ distribution of these 126 objects is shown in Fig. 6 (c). The distribution exhibits a stepwise increase in the number of occurrences as the $G_{\mathrm{mag}}$ increases. This figure reveals that faint objects are susceptible to moonlight contamination.

Overall, we confirmed that $\rho$, $G_{\mathrm{mag}}$, and lunar phase are key factors in determining which spectra are impacted by moonlight contamination. In Fig. 7, we present the 3D scatter plot of these 125 stars, along with the 2D scatter plots of the three directional slices, each accompanied by a colour bar. Analysis shows that, among these 126 stars, there are no instances where the lunar phase is far from the full moon phase, the $\rho$ value is large, and the star's $G$-band magnitude is bright simultaneously.

Next, we plot these 126 objects on the colour–magnitude diagram (CMD). We obtained the 100 pc astrometrically 'clean' subsets from *Gaia* DR3 (Lindegren et al. 2018) to serve as the background in Fig. 8. For these 125 stars, we use `astropy.coordinates.SkyCoord` (Astropy Collaboration 2022) represents their position in the sky. We then applied the `dustmaps` (Green 2018) to query and correct the interstellar extinction. The results are shown in Fig. 8, where the contaminated objects are marked as red triangles. Many of these objects align along the single-star sequence, suggesting they are single stars rather than binaries.

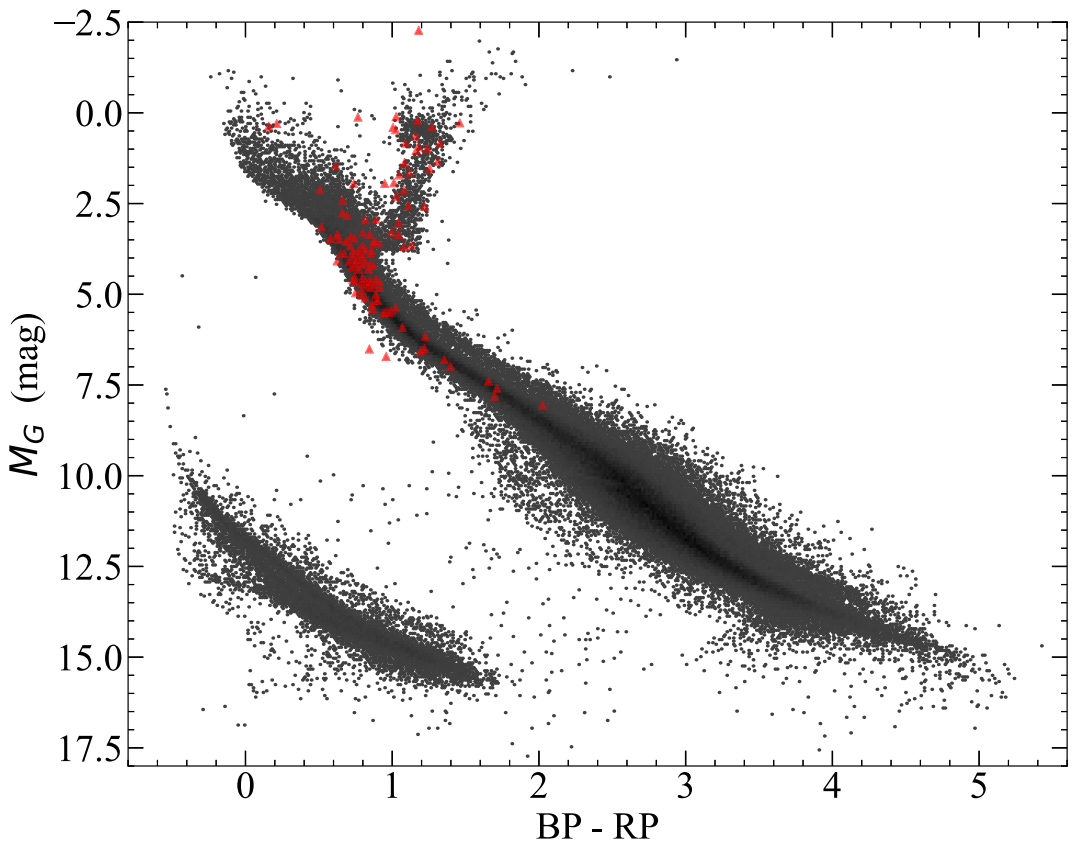

**Figure 8.** This CMD presented with the *Gaia* DR3 astrometrically 'clean' subsets within 100 pc as a grey background. Spectra identified as contaminated are marked with red triangles, and most of them follow the single-star sequence.

## 5 CONCLUSIONS

We successfully detected 126 false-positive SB2 by analysing LAMOST-MRS spectra. These contaminated stars typically exhibit single-star spectral lines with consistent $RV_0$ values across all spectra, except for one polluted spectrum showing double-line characteristics. We have proven that the secondary component is derived from moonlight pollution, as it exhibits near-zero radial velocity and resembles solar-like spectral lines. Additionally, we investigated the correlation between false positive SB2 detections and factors such as the lunar phase, $\rho$, and $G_{\rm mag}$. This helps us understand the mechanisms of moonlight contamination.

We also suspect some contamination may result from low-orbital artificial satellites and stray light. However, this is not the primary focus of the current study, and we have not addressed it in detail here. We plan to acquire satellite launch data and investigate satellite-related contamination.

In future work, we aim to identify additional false positive SB2 systems, classify them according to their contamination sources, and investigate cases where contamination occurs repeatedly across multiple time-domain observations. Stars may be subject to various contamination sources, presenting both an intriguing and challenging area of study. Furthermore, we plan to develop methods for more effectively disentangling contamination components during data processing.

## ACKNOWLEDGEMENTS

This work is supported by the National Natural Science Foundation of China (grants nos. 12288102, 12125303, and 12090040/3), the National Key R&D Program of China (grant no. 2021YFA1600403), the International Centre of Supernovae, Yunnan Key Laboratory (no. 202302AN360001), the Yunnan Revitalization Talent Support Program – Science & Technology Champion Project (grant no. 202305AB350003) and Natural Science Foundation of Yunnan Province (no. 202201BC070003).

## DATA AVAILABILITY

The data underlying this article will be shared at a reasonable request by the corresponding author. LAMOST-MRS spectra are downloaded from www.lamost.org. The solar spectrum is downloaded from https://bass2000.obspm.fr/solar_spect.php.

## REFERENCES

Abt H. A., 1983, ARA&A, 21, 343
Astropy Collaboration, 2022, ApJ, 935, 167
Bassa C., Hainaut O., Galadí-Enríquez D., 2022, A&A, 657, A75
Chen X., Wang S., Deng L., De Grijs R., Yang M., Tian H., 2020, ApJS, 249, 18
Cui X.-Q. et al., 2012, Res. Astron. Astrophys., 12, 1197
Czesla S., Schröter S., Schneider C. P., Huber K. F., Pfeifer F., Andreasen D. T., Zechmeister M., 2019, PyA: Python astronomy-related packages, preprint (ascl:1906.010)
Duquennoy A., Mayor M., 1991, A&A, 248, 485
El-Badry K., Rix H.-W., Ting Y.-S., Weisz D. R., Bergemann M., Cargile P., Conroy C., Eilers A.-C., 2018a, MNRAS, 473, 5043
El-Badry K. et al., 2018b, MNRAS, 476, 528
Fernandez M. et al., 2017, PASP, 129, 084201
Ge H., Webbink R. F., Han Z., Chen X., 2010, Ap&SS, 329, 243
Green G., 2018, J. Open Source Softw., 3, 695
Guo Y., Liu C., Wang L., Wang J., Zhang B., Ji K., Han Z., Chen X., 2022, A&A, 667, A44
Halbwachs J., Mayor M., Udry S., Arenou F., 2003, A&A, 397, 159
Han Z.-W., Ge H.-W., Chen X.-F., Chen H.-L., 2020, Res. Astron. Astrophys., 20, 161
Hilditch R. W., 2001, An Introduction to Close Binary Stars. Cambridge Univ. Press, Cambridge
Holmberg J., Nordström B., Andersen J., 2009, A&A, 501, 941
IJspeert L. W., Tkachenko A., Johnston C., Garcia S., De Ridder J., Van Reeth T., Aerts C., 2021, A&A, 652, A120
Kounkel M. et al., 2021, AJ, 162, 184
Kovalev M., Heidelberg university 2019, PhD thesis
Kovalev M., Chen X., Han Z., 2022, MNRAS, 517, 356
Kovalev M., Zhou Z., Chen X., Han Z., 2024, MNRAS, 527, 521
Li C.-Q., Shi J.-R., Yan H.-L., Fu J.-N., Li J.-D., Hou Y.-H., 2021, ApJS, 256, 31
Lindegren L. et al., 2018, A&A, 616, A2
Liu J. et al., 2019, Nature, 575, 618
Liu C. et al., 2020, preprint (arXiv:2005.07210)
Matijevič G. et al., 2010, AJ, 140, 184
Matijevič G. et al., 2011, AJ, 141, 200
Merle T. et al., 2017, A&A, 608, A95
Moe M., Di Stefano R., 2017, ApJS, 230, 15
Nordström B., Andersen J., Holmberg J., Jørgensen B. R., Mayor M., Pont F., 2004, Publ. Astron. Soc. Aust., 21, 129
Pourbaix D. et al., 2004, A&A, 424, 727
Prša A. et al., 2022, ApJS, 258, 16
Prusti T. et al., 2016, A&A, 595, A1
Raghavan D. et al., 2010, ApJS, 190, 1
Rowan D. et al., 2022, MNRAS, 517, 2190
Rowan D. et al., 2023a, MNRAS, 520, 2386
Rowan D., Jayasinghe T., Stanek K., Kochanek C., Thompson T. A., Shappee B., Giles W., 2023b, MNRAS, 523, 2641
Schneider R., Ferrari V., Matarrese S., Portegies Zwart S. F., 2001, MNRAS, 324, 797
Slawson R. W. et al., 2011, AJ, 142, 160
Traven G. et al., 2017, ApJS, 228, 24
Traven G. et al., 2020, A&A, 638, A145
Vallenari A. et al., 2023, A&A, 674, A1
Zhang B. et al., 2022, ApJS, 258, 26
Zhao G., Zhao Y.-H., Chu Y.-Q., Jing Y.-P., Deng L.-C., 2012, Res. Astron. Astrophys., 12, 723

## APPENDIX A: ANALYSING EXCLUDED DATA

For these 15 stars, if they are indeed SB2 systems, we infer that they likely possess highly eccentric, long-period orbits, which would explain the appearance of double lines characteristic in only one observation. To confirm whether they are contaminated, we analysed their spectra following the method described in the main text and provided the $RV_2 + RV_{\rm h}$ values for 15 potentially contaminated spectra in Table A1.

We carefully examined the spectra of these stars and confirmed that all were contaminated. The sources of contamination are various, including moonlight, low-orbit satellites, stray light, and more. In Table A1, we divide them into three categories: contamination is marked in boldface (6 targets), those with non-moonlight contamination are underlined (6 targets), and cases where the source of contamination cannot be determined (3 targets). For each category, we select one representative star for demonstration, resulting in 3 figures (Fig. A1–A3). The top panels of these figures display all observed spectra for each star. The middle panels display the individual 20-min exposure spectra of the contaminated observation. The bottom panels present the fitting results, reduced chi-square values, and residuals between the observed and fitted spectra.

Fig. A1 illustrates that J062659.14+205958.1 is contaminated by moonlight. MJD of the contaminated spectrum is 59270.512, marked by purple lines in the top panel. The middle panel displays four individual 20-min exposure spectra of this observation, showing distinct double-line features, particularly in the Mgb line. The fitting results in the bottom panel indicate $\gamma$ values of 0.763 and 0.865 for the blue and red arms, respectively, suggesting that this star is cooler than the Sun. The reduced chi-square value, $\chi^2_\nu \sim 0.529$, demonstrates a good fit. For the contaminated spectrum, $RV_2 + RV_{\rm h} \sim 10.25937\ {\rm km \cdot s^{-1}}$, which slightly exceeds the criterion of $|RV_2 + RV_{\rm h}| < 10\ {\rm km \cdot s^{-1}}$. This indicates that the established criteria may be overly strict, potentially excluding some stars affected by contamination.

Fig. A2 illustrates that J103522.75+074710.2 is contaminated, but not by moonlight. MJD of the contaminated spectrum is 59676.566, marked by pink lines in the top panel. The middle panel displays three individual 20-min exposure spectra of this observation, showing distinct double-line features, particularly in the H$\alpha$ and Mgb lines. The bottom panel displays $\chi^2_\nu \sim 10.641$, indicating a poor fit. Moreover, the contaminated spectrum (grey) and fitted spectrum (green) are poorly aligned at the absorption lines. And the $RV_2 + RV_{\rm h}$ value of the contaminated spectrum is $91.85\ {\rm km \cdot s^{-1}}$, which is very large. These factors together suggest that the contamination is not a solar-like spectrum, and this star is likely polluted by non-moonlight sources.

Fig. A3 illustrates that J112651.18+021201.1 is contaminated, but the source of the contamination cannot be determined. MJD of the contaminated spectrum is 59685.608, marked by brown lines in the top panel. The $RV_2 + RV_{\rm h}$ value of the contaminated spectrum is $16.31\ {\rm km \cdot s^{-1}}$, which exceeds the criterion of $|RV_2 + RV_{\rm h}| < 10\ {\rm km \cdot s^{-1}}$. The middle panel displays six individual 20-min exposure spectra of this observation, each showing spectral line profile broadening at the H$\alpha$ line, indicating that this spectrum is contaminated. In the bottom panel, the $\chi^2_\nu$ is 0.923, indicating the fit is not bad. Additionally, the $\gamma$ values for the blue and red arms are 0.693 and 0.764, respectively, indicating this star is cooler than the Sun In general, this star is contaminated; however, given the relatively large $RV_2 + RV_{\rm h}$ value, we cannot conclusively determine whether the source of contamination is moonlight.

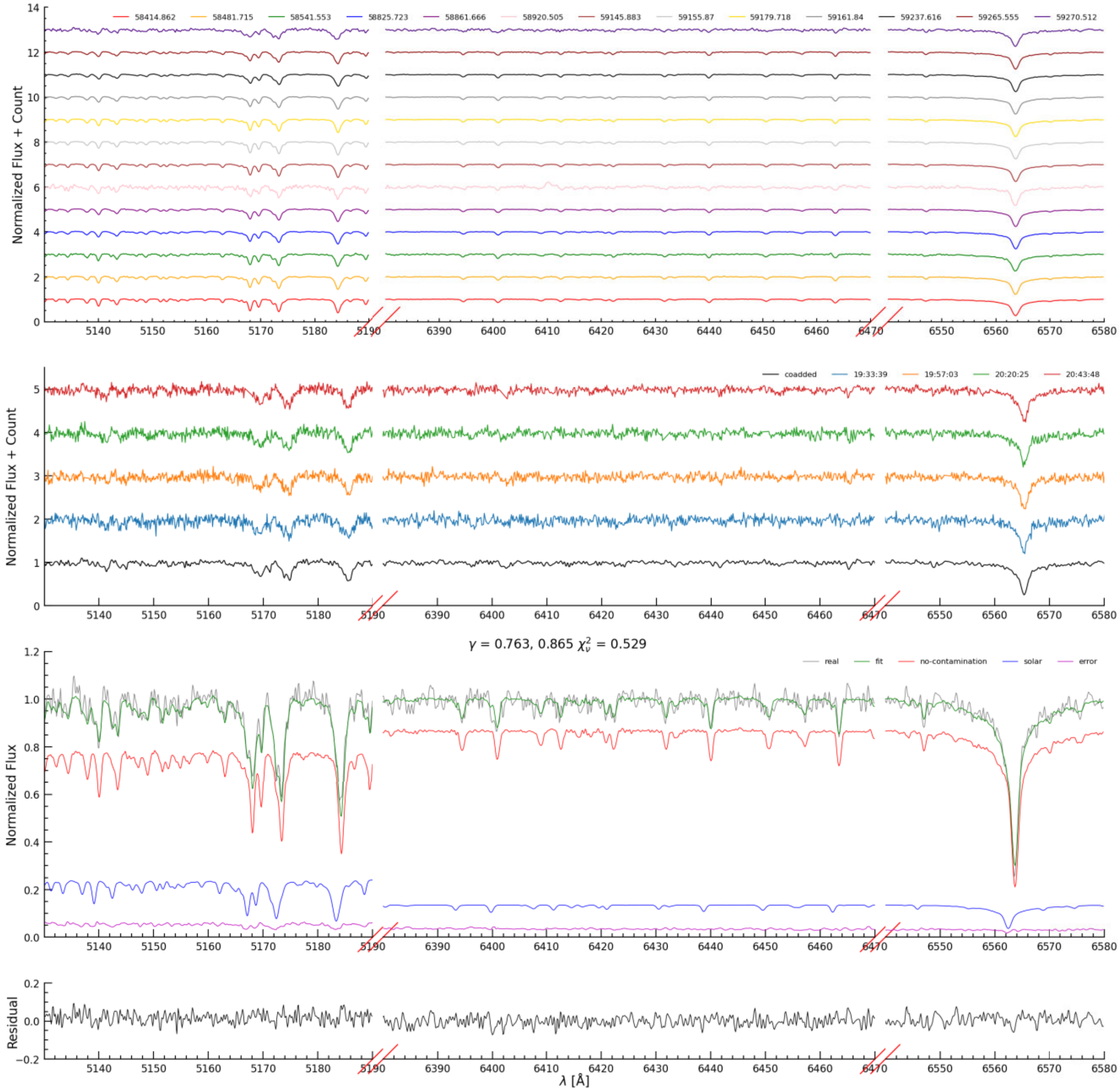

**Figure A1.** J062659.14+205958.1 is contaminated by moonlight. The MJD of the contaminated spectrum is 59270.512 and the MJD of the non-contaminated spectrum used for fitting is 58825.723.

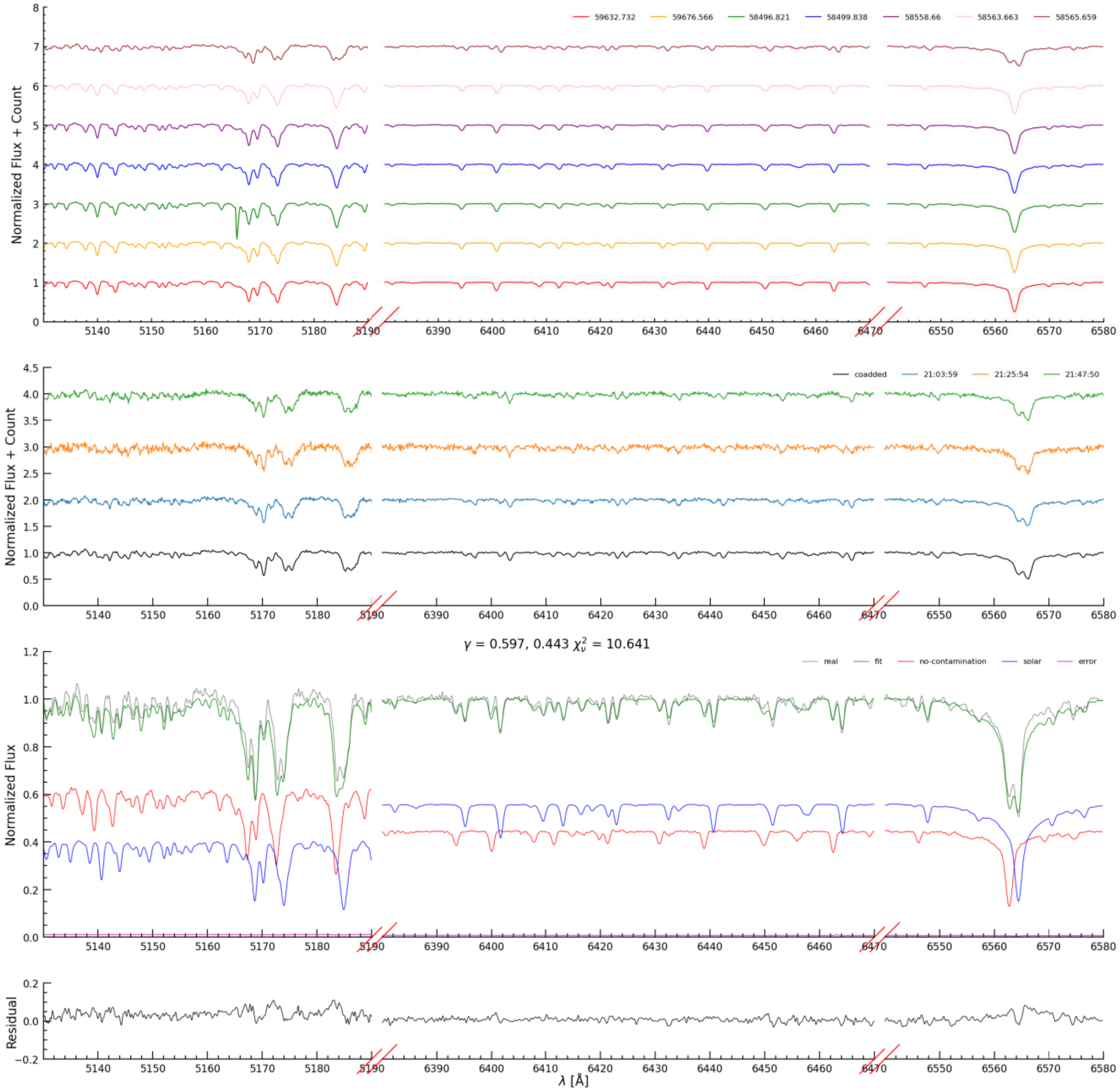

**Figure A2.** J103522.75+074710.2 are contaminated, but not by moonlight. The MJD of the contaminated spectrum is 59676.566, and the MJD of the non-contaminated spectrum used for fitting is 59632.732.

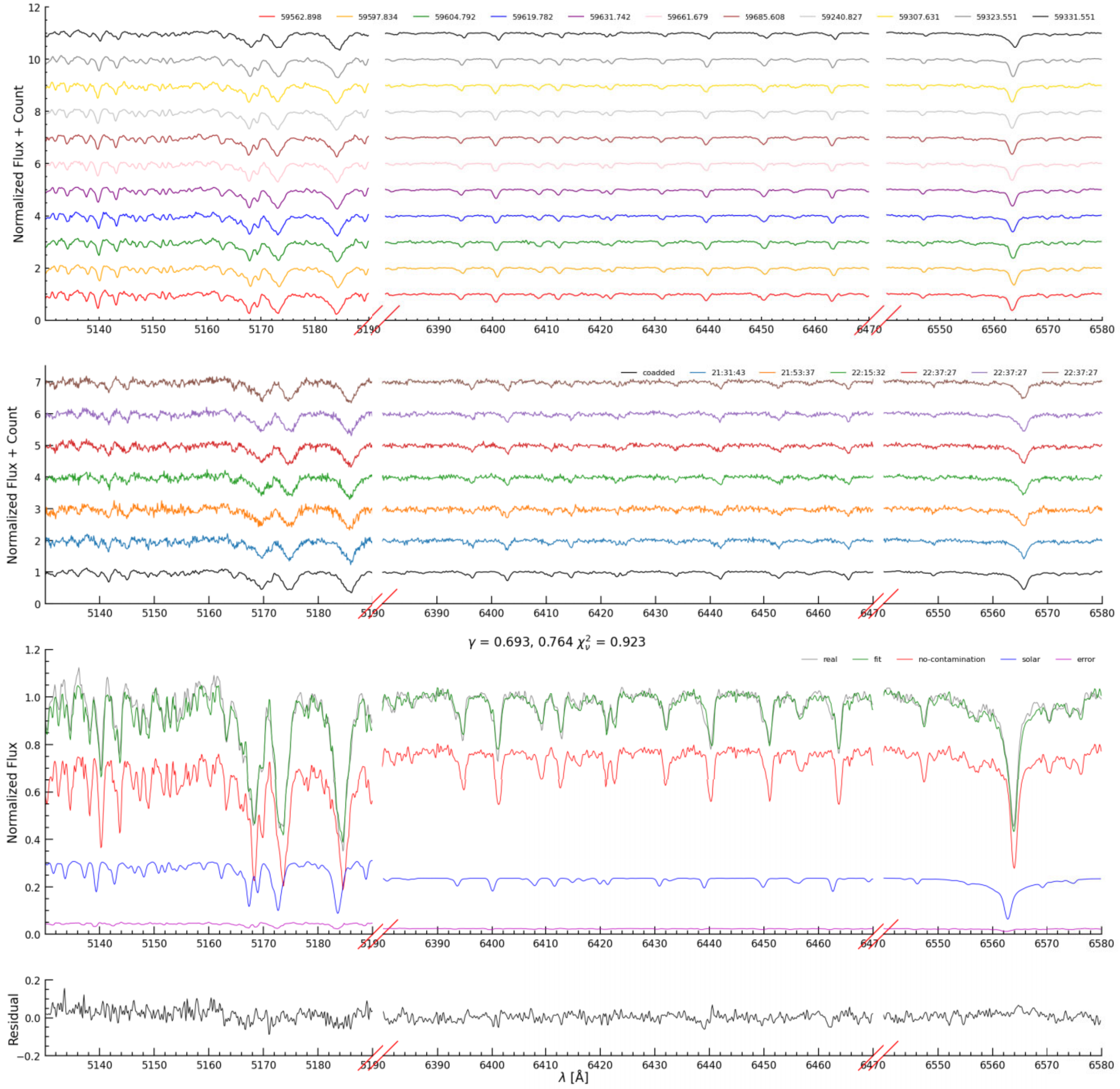

**Figure A3.** J112651.18+021201.1 are contaminated, we cannot determine the source of contamination. The MJD of the contaminated spectrum is 59685.608, and the MJD of the non-contaminated spectrum used for fitting is 59597.834.

**Table A1.** We list Designations and $RV_2 + RV_\mathrm{h}$ values for 15 potentially contaminated spectra, all confirmed to be contaminated. These spectra are categorized into three groups: moonlight contamination (indicated in boldface), non-moonlight contamination (indicated by underline), and cases where the source of contamination remains undetermined.

| Designation | $RV_2 + RV_\mathrm{h}$ km · s$^{-1}$ |
|---|---|
| <u>J151125.28+535115.8</u> | −71.74 |
| J104820.81+551653.0 | −47.76 |
| <u>J060537.13+212950.5</u> | −24.06 |
| **J104753.13+055212.5** | −10.57 |
| **J113242.73+032826.0** | 10.22 |
| **J062659.14+205958.1** | 10.26 |
| **J141248.89+441344.2** | 11.57 |
| **J104727.37+091253.1** | 11.65 |
| **J125828.88+540024.0** | 15.6 |
| J112651.18+021201.1 | 16.31 |
| J092022.05+410642.8 | 21.26 |
| <u>J103235.90+073907.5</u> | 82.18 |
| <u>J141649.20+440404.6</u> | 83.89 |
| <u>J103522.75+074710.2</u> | 91.85 |
| <u>J034647.70+245359.4</u> | 97.77 |

This paper has been typeset from a TEX/LATEX file prepared by the author.